\documentclass[submission,Phys]{SciPost}
\usepackage{amsmath,cases}
\usepackage[utf8]{inputenc}
\DeclareUnicodeCharacter{2215}{/}
\usepackage[T1]{fontenc}
\usepackage{lmodern}
\usepackage{graphicx}
\usepackage{amssymb}
\usepackage{gensymb}
\usepackage{bm}
\usepackage{tabularx}
\usepackage{mathtools}
\usepackage{microtype}
\usepackage{siunitx}
\usepackage[version=4]{mhchem}
\usepackage{bbm}
\usepackage{braket}
\usepackage{mathtools}
\usepackage[normalem]{ulem}
\usepackage{layouts}
\usepackage{float}
\usepackage{textgreek}
\usepackage{orcidlink}

\usepackage{datatool, filecontents}
\DTLsetseparator{,}

\hypersetup{colorlinks=true, linkcolor=blue, citecolor=blue, urlcolor=blue}

\newcounter{CommentNumber}
\renewcommand{\paragraph}[1]{\stepcounter{CommentNumber}\belowpdfbookmark{#1}{\arabic{CommentNumber}}}

\begin{document}

\begin{center}{\Large \textbf{
 Probing valley phenomena with gate-defined valley splitters
}}\end{center}

\begin{center}
 Juan Daniel Torres Luna\textsuperscript{1, 2, $\dag$}\orcidlink{0009-0001-6542-6518},
 Kostas Vilkelis\textsuperscript{1, 2, $\ddag$}\orcidlink{0000-0003-3372-1018},
 Antonio L. R. Manesco\textsuperscript{2, *}\orcidlink{0000-0001-7667-6283}.
\end{center}
\begin{center}
 {\bf 1} Qutech, Delft University of Technology, Delft 2600 GA, The Netherlands
    \\
 {\bf 2} Kavli Institute of Nanoscience, Delft University of Technology, Delft 2600 GA, The Netherlands
    \\
    \textsuperscript{$\dag$}jd.torres1595@gmail.com, \textsuperscript{$\ddag$}kostasvilkelis@gmail.com, \textsuperscript{*}am@antoniomanesco.org
\end{center}
\begin{center}
 May 1, 2024
\end{center}

\section*{Abstract}

Despite many reports of valley-related phenomena in graphene and its multilayers, current transport experiments cannot probe valley phenomena without the application of external fields.
Here we propose a gate-defined valley splitter as a direct transport probe for valley phenomenon in graphene multilayers.
First, we show how the device works, its magnetotransport response, and its robustness against fabrication errors.
Secondly, we present two applications for valley splitters: (i) resonant tunneling of quantum dots probed by a valley splitter shows the valley polarization of dot levels; (ii) a combination of two valley splitters resolves the nature of order parameters in mesoscopic samples.

\section{Introduction}

\paragraph{Recent progress in fabrication made valley phenomena experimentally accessible.}

The recent advances in the fabrication of graphene multilayer devices, particularly boron nitride encapsulation~\cite{Dean2010, Dean2012} and the use of graphite back gates~\cite{Standley2012, Eich2018, Banszerus2018}, improved the mobility and suppressed intervalley scattering of bulk samples.
Due to valley conservation in this new generation of clean bulk graphene samples, many experimental works observed valley order arising from spontaneous $SU(4)$ spin-valley symmetry breaking~\cite{Zhou2022, Huang2022, delaBarrera2022,Cao2018,Chatterjee2022,Nuckolls2023}.
In addition, the out-of-plane displacement field opens a gap in graphene multilayers~\cite{Mak2009, Castro2007, Oostinga2007, Zhang2009}, and therefore allows to fabricate gate-defined nanodevices~\cite{Allen2012, Overweg2017, Goossens2012, Eich2018}.
The smooth potential of electrostatically-defined devices avoids intervalley scattering caused by reflection at atomically-sharp edges.
As a consequence, valley-dependent phenomena were also observed in nanostructures, such as the injection of valley-polarized states in quantum point contacts~\cite{Overweg2017,Sakanashi2021,Gold2021,InglaAyns2023,Ingla2023}, and valley-polarized in states quantum dots~\cite{Eich2018,Banszerus2018} with long lifetimes~\cite{Garreis2024,Banszerus2024_phononT1}.

\paragraph{Despite the recent observations, their measurements are often indirect and require external magnetic fields.}

The recent experiments in graphene multilayer devices observe various symmetry-broken phases that suggest the presence of valley structure, such as nematic superconductivity~\cite{Cao2018}, isospin ferromagnetism~\cite{Chatterjee2022}, and correlated insulators~\cite{Nuckolls2023}.
However, current transport experiments cannot resolve the valley structure of the order parameter.
For example, quantum oscillation experiments cannot distinguish between valley polarized and intervalley coherent (IVC) orders.
In nanodevices, experiments probing valley-related phenomena rely on breaking valley symmetry and therefore these probes are invasive and indirect.
For example, valley polarization of states injected from quantum point contacts was probed with electron collimation~\cite{Ingla2023} and via valley-Zeeman splitting in quantum dots~\cite{Tong2021, Lee2020}.
Finally, transport signatures of valley polarization using Andreev reflection~\cite{Titov2006,Akhmerov2007,Beenakker2008,Niu2019} are still limited by device quality~\cite{Zhao2020,Manesco2022,Bhandari2020}.
Thus, the only direct probes of valley phenomena are via microscopy techniques~\cite {Zhang2020,Liu2022,Gold2021,Nuckolls2023}.

\paragraph{We explore the use of gate-defined valley splitters to probe valley phenomenon in BLG.}

In this manuscript, we propose the use of electrically defined valley splitters in bilayer graphene as a magnetic field-free transport probe for valley phenomena.
Although the ingredients of this device were previously explored both theoretically~\cite{Martin2008,Jung2011,FabianThomas2021,PrikoszovichKonrad2022} and experimentally~\cite{Li2016,Kang2018,Li2018,PhysRevApplied.11.044033,chen2020gate,Huang_2024,davydov2024easy}, previous works focused on the generation of valley-polarized currents rather than its use as an experimental probe of valley polarization.
We demonstrate the use of valley splitters in two case uses: (i) a tool for measuring valley polarization of confined states via resonant tunneling, and (ii) a probe of valley-dependent order parameters.
We also evaluate the robustness of the splitter's efficiency against fabrication imperfections, show the requirements for the alignment of splitting gates, and suggest changes in the device layout to further relax the conditions.

\section{Device}

\begin{figure}[!b]
    \centering
    \includegraphics[width=0.95\textwidth]{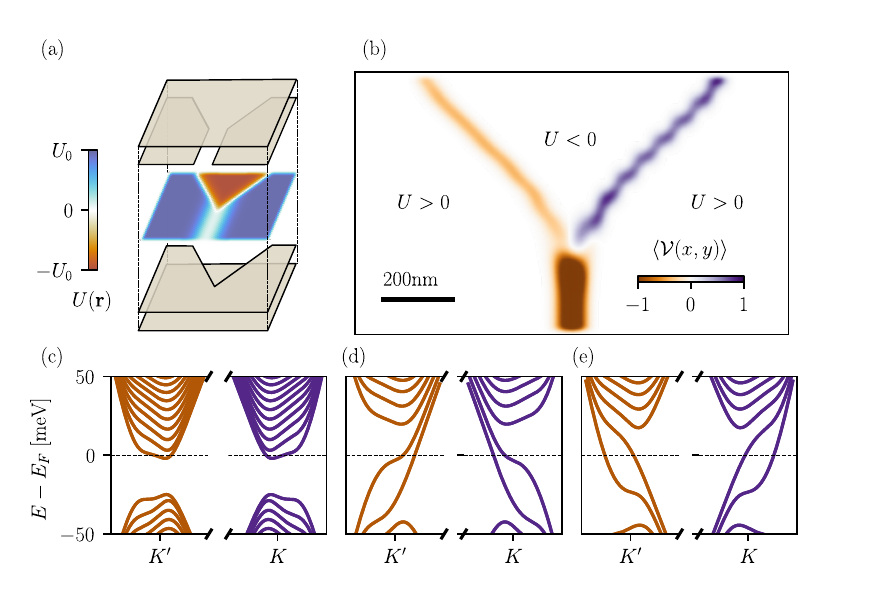}
    \caption{
 Tight-binding simulations of a valley splitter.
 (a) Proposed multilayered gate layout and resulting displacement field in bilayer graphene.
 (b) Valley expectation value of the scattering wave functions.
 The injecting states are superpositions of states at both valleys.
 As they arrive at the helical edge channels, each flavor leaves through one of the collector arms.
 The difference between transmissions from upper and lower leads is proportional to the valley polarization of the injected modes.
 (c) Band structure of the injecting one-dimensional channel.
 The presence of inversion and time-reversal symmetry results in a valley-symmetric band structure.
 Band structure of the (d) left and (e) right collecting channels.
 The highlighted bands are valley-helical, resulting in the splitting effect as injected modes encounter the collecting arms.
 }
    \label{fig:device}
\end{figure}

\paragraph{We propose a valley splitter device which consists of a one-dimensional injection channel and two valley-helical collector arms}

We propose a bilayer graphene gate-defined device depicted in Fig.~\ref{fig:device}(a), that splits states according to their valley polarization.
The devices consist of three one-dimensional channels confined by gapped regions due to a non-zero displacement field $D(\mathbf{r})$ as shown in Fig.~\ref{fig:device}(b).
The injection channel is confined by a region with $D>0$ and injects valley-symmetric states.
The collecting arms are valley-helical channels surrounded by regions with opposite displacement field~\cite{Martin2008,Jung2011, Li2016}.
Because of the opposite sign of the displacement field, one arm collects states at valley $K$ and the other arm collects states at the opposite valley $K'$.
We propose the device layout from Fig.~\ref{fig:device}(a) to achieve the required displacement field profile.
The control of the four gates provides independent control of the layer imbalance $U\propto D$ and the chemical potential $\mu$.

\paragraph{We study the valley splitter using tight-binding simulations.}

To demonstrate the transport properties of the valley splitter, we perform numerical simulations using Kwant~\cite{Groth2014}.
We implement the Slonczewski-Weiss-McClure parametrization~\cite{PhysRev.108.612,PhysRev.109.272,PhysRev.119.606,McCann2013} (see Fig.\ref{fig:scaling} (a) for details)
and using tight-binding parameters obtained via infrared spectroscopy~\cite{PhysRevB.80.165406}:
\begin{align}
 \mathcal{H} = \sum_{n} \psi_n^{\dagger}\left[U (\mathbf{r}_n) \mathrm{sign}(\mathbf{r}_n\cdot \hat{z}) - \mu(\mathbf{r}_n) + g\mu_B B\sigma_z \right]\psi_{n} + \sum_{\left\lbrace n, m \right\rbrace \in S_i} \gamma_i e^{i\phi_{nm}} \psi_{n}^{\dagger}\psi_{m},
    \label{eq:hamiltonian}
\end{align}
where $\psi_n = (c_{n\uparrow}, c_{n\downarrow})^T$,  $c_{n, \sigma}$ is an annihilation operator of an electron with spin $\sigma$ at the atomic site $n$.
The matrices $\sigma_i$ are Pauli matrices that act on spin degrees of freedom.
Here $\mu(\mathbf{r})$ and $U(\mathbf{r})$ are the position-dependent chemical potential and layer imbalance.
To exclude the effects of edge states, we add an staggered potential at the device's boundary.
Furthermore, $g\approx 2$ is the $g$-factor in bilayer graphene, $\mu_B$ is the Bohr magneton, and $B$ is the out-of-plane magnetic field.
Due to the magnetic field $B$, the hoppings between sites $n$ and $m$ are corrected by a Peierls phase $\phi_{nm}$.
The hopping energies are set according to Ref.~\cite{McCann2013}: each hopping energy $\gamma_i$ corresponds to a set $S_i$ of hopping vectors between the site pairs $(n, m)$.
The electrostatic potential set by $\mu(\mathbf{r})$ and $U(\mathbf{r})$ has a screening length of $\SI{60}{\nano\metre}$~\cite{Flr2022,Li2020}.\footnote{To avoid the complexity of electrostatic simulations, we pass a gaussian filter with $\sigma=\SI{30}{\nano\metre}$ to step-functions that define the electrostatic potential according to the layout shown in Fig.~\ref{fig:device}(a).}
To reduce the computational cost of the calculations, we rescale the in-plane lattice constant to $\tilde{a} = sa$ by a factor $s=16$ and fix the hoppings accordingly to preserve the low-energy dispersion.
The scaled model and its validity is discussed in the Appendix~\ref{app:scaling}.

\begin{figure}[h]
    \centering
    \includegraphics[width=0.95\textwidth]{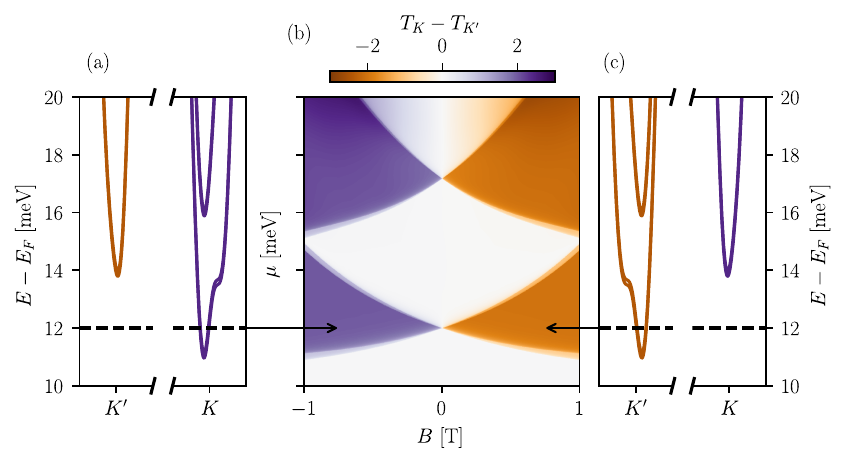}
    \caption{
 Magnetotransport simulations with a valley splitter.
 (a, c) Band structure of the injecting arms at finite orbital field $B$ which lifts the valley degeneracy. 
 As a consequence, the number of propagating modes per valley changes.
 (c) Transmission difference between the two helical collecting arms.
 A finite difference indicates the valley polarization of the injected modes.
 The transmission changes in integer steps, proportionally to the net number of propagating modes per valley.
 }
\label{fig:transport}
\end{figure}

\paragraph{We confirm the splitter behaviour with tight-binding simulations.}
We show the tight-binding simulation for the valley splitter device in Fig.~\ref{fig:device}.
In all simulations in this manuscript, we set the distance of $\SI{100}{\nano\metre}$ between split gates in the injecting channel.
The layer imbalance away from the one-dimensional channels is fixed at $U=\pm 100$ meV.
States are injected in the device through the narrow channel at the bottom of Fig.~\ref{fig:device}(b).
The bandstrcuture in this channel is shown in Fig.~\ref{fig:device}(c), which is valley symmetric and therefore allows equal injection of states in both valleys.
As states propagate, they encounter two arms at the top of Fig.~\ref{fig:device}(b).
Because of the displacement field configuration, the arms have opposite valley helicity, as shown in the bandstructure plots of Figs.~\ref{fig:device}(d, e).
The scattering wavefunctions $\psi$ are spatially split according to their valley polarization, shown by their valley expectation value $\langle \psi | \mathcal{V} | \psi \rangle$ in Fig.~\ref{fig:device}(b).\footnote{The valley operator $\mathcal{V}$ is defined as a sublattice-dependent Haldane coupling~\cite{Coloms2018,Ramires2018,Manesco2022}.}
The smooth electrostatic potential ensures adiabatic propagation of wavefunctions and prevents both backscattering and valley mixing.
As a result, each injected mode is perfectly transferred to the lead with the matching valley helicity.

\paragraph{We show that the transmission difference in the valley splitter measures the valley polarization of injected modes.}

The valley splitter directly probes the valley polarization of the injected current.
We define the net valley polarization of a set of modes $\Phi_i$ as
\begin{equation}
 V = \frac{\sum_i \langle \Phi_i |\mathcal{V} | \Phi_i \rangle}{\sum_i \langle \Phi_i | \Phi_i \rangle}~.
\end{equation}
Moreover, defining the transmission to arms that collect states in the valley $K$ and $K^\prime$ as $T_K$ and $T_{K^\prime}$ and assuming a splitter with perfect efficiency and adiabatic,
\begin{equation}
 T_{\alpha} = \sum_i \langle \Phi_i | P_{\mathcal{\alpha}} | \Phi_i\rangle~,
\end{equation}
where $P_{\mathcal{\alpha}}$ is the projector for the valley $\alpha$.
Thus,
\begin{equation}
 T_V = \frac{T_K - T_{K^\prime}}{T_K + T_{K^\prime}} \equiv V
\end{equation}
as long as the collecting arms are not saturated, $\emph{i.e.}$ $T_{\alpha}$ is smaller than the number of valley-helical channels (2 for bilayer graphene).

\paragraph{We demonstrate the device operation with a magnetotransport calculation.}

We demonstrate the valley-splitter opearation simulating a magnetotransport experiment.
Because an orbital field $B$ lifts the valley degeneracy~\cite{Lee2020,Overweg2018,PhysRevB.101.085118,PhysRevB.98.155435,Tong2021}, we observe a field dependence of the valley transmission.
We show the bandstructure of the injection channel in Figs.~\ref{fig:transport}(a, c), and the resulting valley transmission $T_V$ in Fig.~\ref{fig:transport}(b).
The imbalance in the number of modes per valley $n_{\alpha}$ in the injector results in $T_V = (n_K - n_{K^\prime}) / (n_K + n_{K^\prime})$.
We show the break of valley degeneracy at the injector in Fig.~\ref{fig:transport}(a) at $B < 0$ and Fig.~\ref{fig:transport}(b) at $B > 0$.
The number of modes per valley at the Fermi level in the helical channels does not change with the magnetic field.
Therefore, the difference in transmission between the two arms is proportional to the valley polarization of the injected modes, as demonstrated in Fig.~\ref{fig:transport} (b).
It is important to notice that the equivalence between $T_V$ and $V$ does not require valley conservation, and we explore the generality of our result in Sec.~\ref{sec:applications}.

\paragraph{The efficiency of the splitter depends on the geometry of the device.}

\begin{figure}[ht!]
    \centering
    \includegraphics[width=0.95\textwidth]{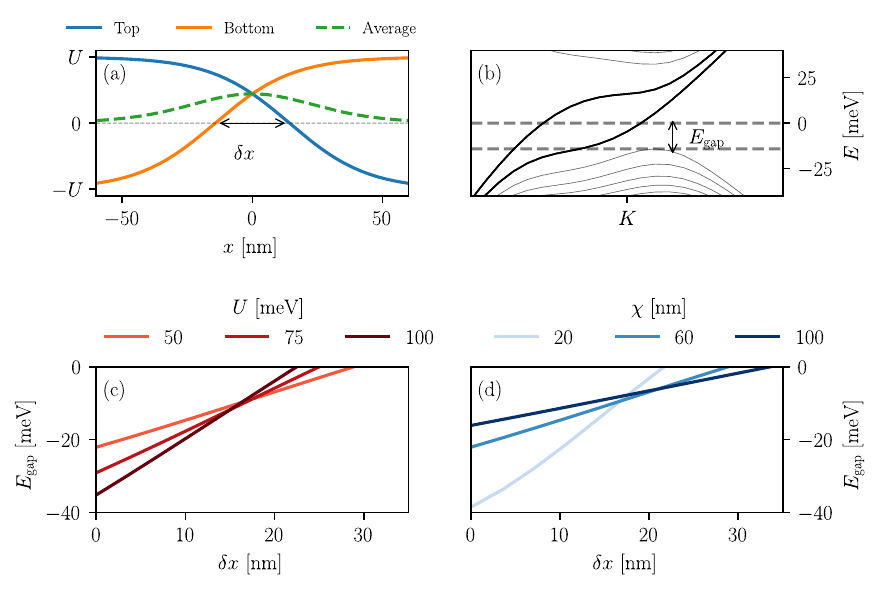}
    \caption{
 Misalignment effects on the valley helical channel.
 (a) Example electrostatic potential used in the layers.
 The solid lines show the potential in each layer.
 The dashed line shows the average potential, which corresponds to a local modulation of the chemical potential.
 (b) The band structure of the valley helical channel at $\delta x = \SI{10}{\nano\metre}$ and $U = \SI{50}{\milli\eV}$.
 Because of the misalignment $\delta x$ and the corresponding modulation of the chemical potential, non-helical bands approach the Fermi energy.
 (c) The bandwidth of the valley helical states as a function of misalignment $\delta{x}$ and layer imbalance $U$.
 (d) The bandwidth of the valley helical states as a function of $\delta x$ and screening length $\chi$ with $U = \SI{50}{\milli\eV}$.
 Plots (a, c, d) use $\chi = \SI{60}{\nano\metre}$.
 }
    \label{fig:misalingmnet}
\end{figure}

Because one of the main fabrication challenges is the alignment between gates, we simulate the effects of misalignments on the valley helical collecting channels in Fig.~\ref{fig:misalingmnet}.
A gate misalignment $\delta{x}$ changes the chemical potential of the helical channels in Fig.~\ref{fig:misalingmnet}(a). 
As a consequence, we observe a shift of the bulk bandstructure, as shown in Fig.~\ref{fig:misalingmnet}(b).
This shift reduces the helical gap, thus reducing the operational window of the valley splitter.
We define the helical gap as the smallest energy window between bulk bands and the Fermi energy $E_{\text{gap}}$, indicated in Fig~\ref{fig:misalingmnet}(b).
Whenever $E_{\text{gap}} > 0$, the valley helical states are outside cross the Fermi energy, and the valley splitter efficiency is suppressed.
In Fig~\ref{fig:misalingmnet}(c,d) we show the dependence of $_{\text{gap}}$ as a function of the displacement field $U$ and the screenng length $\chi$.
Based on Fig.~\ref{fig:misalingmnet}(c), we expect the valley splitter to tolerate misalignments of $20-30 \si{\nano\metre}$, with a higher tolerance for smaller displacement fields.
In addition, the tolerance to misalignments is sensitive to screening length $\chi$: $E_{\text{gap}}$ increases with the screening length as shown in Fig.~\ref{fig:misalingmnet}(d).
It is possible to increase tolerance against gate misalignment with additional gates to control to compensate for the local modulation of the chemical potential, as recently shown experimentally~\cite{Huang_2024}.
Alternatively, a recent work has shown how to fabricate misalignment-robust helical channels~\cite{davydov2024easy}.
We confirm through numerical simulations that the valley splitter is insensitive to the angle between its arms, as discussed in the Appendix~\ref{app:angle}.

\section{Applications}
\label{sec:applications}

\subsection{Resonant tunneling}

\begin{figure}[ht!]
    \centering
    \includegraphics[width=0.95\textwidth]{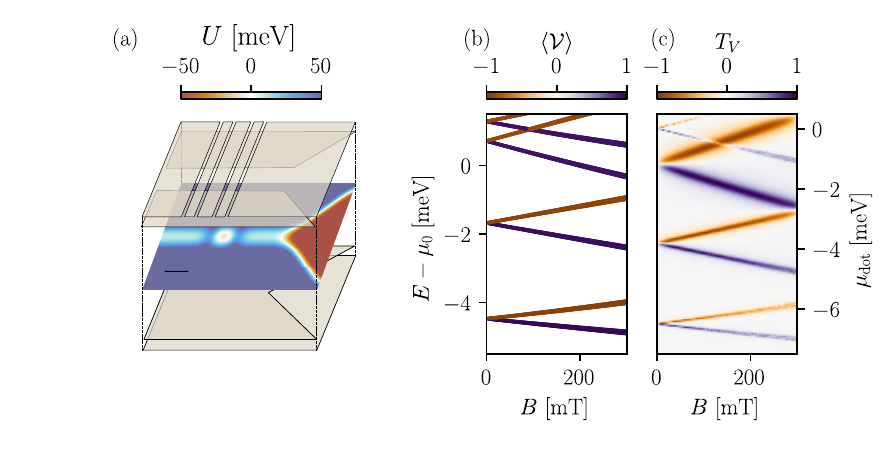}
    \caption{
 Resonant tunnelling through a valley-polarized quantum dot.
 (a) Device schematics showing the gate layout and electrostatic potential of a quantum dot connected to the valley splitter.
 (b) The energy levels of an isolated quantum dot as a function of an out-of-plane magnetic field.
 The colour indicates the valley-polarisation of the states.
 (c) $T_V$ across a quantum dot as a function of an out-of-plane magnetic field and chemical potential $\mu_{\mathrm{dot}}$.
 }
\label{fig:dot}
\end{figure}

\paragraph{We probe the valley polarization of a quantum dot directly with a valley splitter.}

An immediate application of a valley splitter is to probe the valley polarization of quantum dot levels.
We illustrate the operation of the device with numerical simulations of resonant tunnelling through a quantum dot connected to a valley splitter as shown in Fig.~\ref{fig:dot} (a).
To define the quantum dot, we add two additional layer imbalance $t_{\mathrm{dot}}$ regions that provide confinement and also control the tunnelling rate.
For comparison, we show in Fig.~~\ref{fig:dot}(b) the spectrum and corresponding valley polarization of states in an isolated dot as a function of the magnetic field.
Previous experiments in resonant tunneling observe the valley-degeneracy breaking as a function of magnetic field, but cannot resolve the valley polarization of quantum dot states~\cite{Tong2021,Lee2020,Garreis2024,Banszerus2021_spin_valley_dots}.
In Fig.~\ref{fig:dot} (c), we show that connecting the quantum dot to a valley splitter provides full resolution of valley polarization in the quantum dots.
Although we only show data for pristine quantum dots, our setup allows probing valley mixing (see Sec.~\ref{sec:ivc}), which is relevant for driving recent qubit proposals~\cite{denisov2024ultralongrelaxationkramersqubit}.
The small mismatch between the transport experiment and the energy levels of a dot is due to the weak coupling between the dot levels and the leads.

\subsection{Probing valley-dependent order parameter}
\label{sec:ivc}

\begin{figure}[h]
    \centering
    \includegraphics[width=0.95\textwidth]{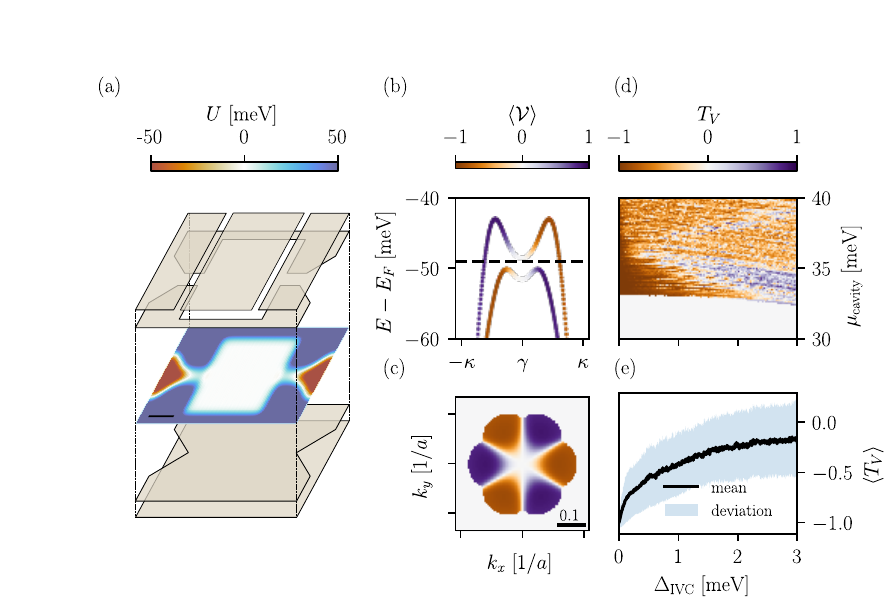}
    \caption{
 A cavity with IVC probed by a double valley splitter setup.
 (a) Gate layout and displacement field in a cavity connected to two valley splitters.
 (b) Band structure of the hole bands in bulk bilayer graphene with IVC.
 (c) Valley texture of the hole pocket for a Fermi energy indicated by the dashed line in panel (b).
 (d) Valley splitter's transmission signal $T_V$ as a function of the cavity's chemical potential and magnitude of the IVC order parameter.
 (e) Average and standard deviation of $T_V$ over the range of $\mu_{\mathrm{cavity}}$ shown in panel (d).
 }
    \label{fig:ivc}
\end{figure}

\paragraph{We propose a device to probe valley-dependent order parameters and demonstrate with IVC.}

Recent experiments showed a variety of electriaccly-tunable symmetry-broken phases in BLG, including valley-ordered phases~\cite{Zhou2022}.
We propose the setup shown in Fig.~\ref{fig:ivc} (a) as an experimental probe of valley order found in these experiments.
The setup consists of a cavity in a valley-symmetry broken phase connected to two valley splitters: one that injects valley-polarized current, and the other probes the valley polarization.
As polarized electrons travel through the cavity, their valley isospin rotates according to the order parameter.
The current at the collecting valley splitter directly relates to the rotation electrons undergo in the cavity.

\paragraph{We exemplify with IVC because it's hard to measure with transport.}

As an example, we demonstrate its use by probing a mesoscopic cavity with intervalley coherent (IVC) order parameter.
In other words, we add to the cavity an order parameter that coherently couples the two valleys~\cite{PhysRevX.10.031034,Xie2023,koh2024}.
As an example of this device operation, we demonstrate its use by probing a mesoscopic cavity with intervalley coherent (IVC) order parameter.
In other words, we add to the cavity an order parameter that coherent couples the two valleys~\cite{PhysRevX.10.031034,Xie2023,koh2024}.
Currently, probes of IVC rely on local probes~\cite{Nuckolls2023,Liu2022,Arp2024,kim2023imaging}.
Thus, probing IVC is impossible in double-gated devices.
On the other hand, transport probes such as quantum oscillations cannot distinguish an intervalley coherent (states at the equator of the valley Bloch sphere) from a valley polarized (states at the poles of the valley Bloch sphere).
Moreover, transport probes such as quantum oscillations cannot distinguish a intervalley coherent (states at the equator of the valley Bloch sphere) from a valley polarized (states at the poles of the valley Bloch sphere).

\paragraph{We simulate the cavity with a Kekulé onsite modulation to induce IVC}

We add an IVC order parameter as a Kekulé onsite modulation within the cavity~\cite{Nuckolls2023,PhysRevLett.131.146601}:
\begin{equation}
 \mathcal{H}_{\mathrm{IVC}} = \Delta_{\mathrm{IVC}} \sum_{m}\sum_{n \in \lbrace A2, B1 \rbrace} \cos\left(\boldsymbol{\eta}_m \cdot \mathbf{r}_n \right)
\end{equation}
where $\boldsymbol{\eta}_m = \Delta K (\cos(2m\pi / 3), \sin(2m\pi / 3))$, and $\Delta K$ is the momentum separation between valleys.
The modulation only acts on the sublattices $A2$ and $B1$ (see Fig.~\ref{fig:scaling}(a)) since these sites set the low-energy manifold~\cite{McCann2013,PhysRevLett.131.146601}.
We show in Fig.~\ref{fig:ivc}(b) the valance bands bulk region with Kekulé modulation.
The modulation mixes valleys and results in avoided crossings in the band structure.
In Fig.~\ref{fig:ivc} we show the $k$-space valley texture of a hole pocket in Fig.~\ref{fig:ivc}(c), consistent with previous works~\cite{Xie2023,Chatterjee2022,PhysRevLett.131.146601}.

\paragraph{We show that valley mixing increases with $\Delta_{\mathrm{IVC}}$.}

We show the effects of the IVC cavity on the valley splitter's transmission signal $T_V$ in the upper panel of Fig.~\ref{fig:ivc}(d).
In the absence of intervalley coherence, $\Delta_{\mathrm{IVC}}=0$, the quasiparticles collected by the splitter transmit only to the arm with the same valley polarization as the injected state, thus $T_V = V_{\mathrm{injection}}$.
As $\Delta_{\mathrm{IVC}}$ increases, we observe that $T_V$ fluctuates due to the valley-dependent dynamical phase accumulated by the quasiparticles in the cavity.
Furthermore averaging over the cavity's chemical potential $\mu_{\mathrm{cavity}}$ (Fig.~\ref{fig:ivc}(e)) we observe a monotonic increase of transmission to the channel with opposite valley helicity.
Therefore, we conclude that $T_V \neq V_{\mathrm{injection}}$ indicates valley mixing in the cavity.

\paragraph{Changing temperature and electrostatics allows us to distinguish between internal coherence and intervalley scattering in an experiment.}

At zero temperature, our scheme cannot distinguish between intervalley coherence (caused by a spontaneous symmetry breaking) and intervalley scattering (caused by short-range disorder).
Therefore, we rely on temperature dependence to distinguish between the two.
Above the critical temperature of the IVC phase, the cavity loses its valley polarization whereas short-range scattering remains.
Thus the absence of valley mixing above the critical temperature would rule out intervalley scattering.
Alternatively, it is possible to control phase transitions electrostatically~\cite{Zhou2022}.
Thus, controlling the correlated phases in the cavity also allows us to verify the amount of valley mixing in different phases and check for the presence of intervalley scattering.
These two checks allow us to unambiguously detect IVC orders.

\section{Conclusion}

We proposed a gate-defined valley splitter as a transport probe of valley phenomena in graphene multilayers.
Our layout tolerates gate misalignments $\lesssim \SI{30}{\nano\metre}$, and we suggest directions for engineering the device layout to relax the misalignment threshold even further.
Moreover, we proposed two applications of valley splitters relevant to recent experimental progress in the field.
Due to the correspondence between valley polarization and normalized net transmission through the valley splitter arms, we show that resonant tunnelling in gate-defined quantum dot probes directly the polarization of the dot levels.
Moreover, we show that a combination of two valley splitters -- one to inject valley-polarized states and the other to probe valley polarization -- allows us to distinguish between valley-polarized and intervalley coherent order parameters in mesoscopic cavities.

\section*{Author contributions}

J.T. proposed to investigate gate-defined valley splitters. A.M. and J.T. proposed possible applications. All authors conceived the layout of the devices, performed numerical simulations and prepared the manuscript. A.M. supervised the project.

\section*{Acknowledgements}

The authors acknowledge the inputs of Anton Akhmerov, Johanna Zijderveld, Timo Hyart and Jose Lado, Josep Ingla-Aynés, Luca Banszerus, Valla Fatemi and Andrea Young for useful discussions on theoretical and experimental considerations.
We thank Anton Akhmerov, Johanna Zijderveld, and Isidora Araya Day for feedback on the manuscript.

\subsection*{Funding information}

The project received funding from the European Research Council (ERC) under the European Union’s Horizon 2020 research and innovation program grant agreement No. 828948 (AndQC).

\section*{Data availablity}

All code used in the manuscript is available on Zenodo~\cite{zenodo}.

\bibliography{biblio.bib}

\newpage

\appendix

\section{Scaled model in a honeycomb structure}
\label{app:scaling}

\begin{figure}[h]
    \centering
    \includegraphics[width=\textwidth]{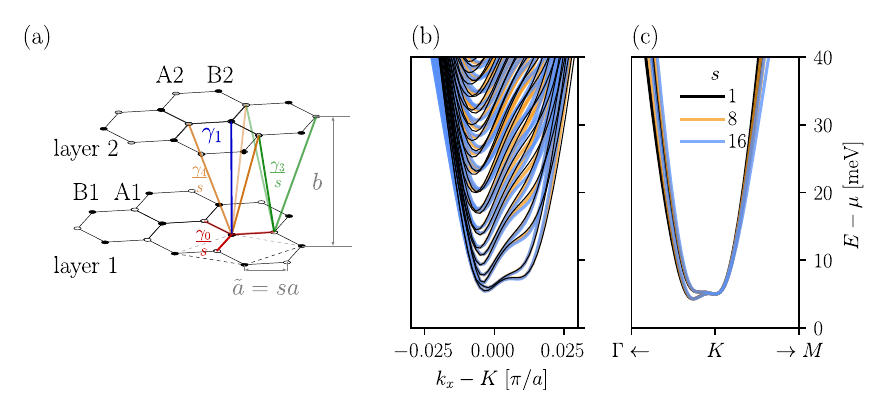}
    \caption{
        (a) Tight-binding model following the Slonczewski-Weiss-McClure parametrization.
        The scaling of the lattice constant $a \mapsto \tilde{a} = sa$ follows a correction $\gamma_i \mapsto \tilde{\gamma}_i = \gamma_i / s$ of the hoppings with in-plane component.
        Band structure of a normal quantum point contact (a) and bulk bilayer graphene (b) around the $K$ point for different scaling factors $s$.
        We show a range of $E - \mu$ used in our transport calculation.
        The quantum point contact in panel (b) has the same parameters as Fig.~\ref{fig:device}(a).
        We add a layer imbalance the bulk calculations shown in (c) of $U = \SI{5}{\milli\electronvolt}$ to match the gap of the point contact for comparison.
    }\label{fig:scaling}
\end{figure}

The simulation of micrometer-sized devices is computationally demanding.
Thus, we rescale the Slonczewski-Weiss-McClure tight-binding model illustrated in Fig.~\ref{fig:scaling}(a) following a procedure inspired by Ref.~\cite{PhysRevLett.114.036601}.
While other discretization schemes have been proposed for Bernal bilayer graphene~\cite{Chen_2024}, our procedure preserves the lattice structure and trigonal warping effects.
The goal of our rescaling scheme is to preserve the low-energy Hamiltonian.
Thus, all hoppings with in-plane components are rescaled as $\gamma_i \mapsto \gamma_i / s$ whereas the only purely out-of-plane hopping $\gamma_1$ is kept as is.

We verify that our scaling scheme preserves the low-energy dispersion comparing the dispersion of a quantum point contact (Fig.~\ref{fig:scaling}(b)) and the bulk dispersion (Fig.~\ref{fig:scaling}(b)) for different scaling factors $s$.
In our transport simulations, we choose $s=16$, resulting in $\tilde{a}=16a\approx\SI{2.3}{\nano\metre}$.
Although we choose $s$ considerably larger than similar works, our choice is well-justified by a simple comparison between length scales, similar to the description in Ref.~\cite{PhysRevLett.114.036601}.
The magnetic length $l_B =\sqrt{\hbar / eB} \approx \SI{25}{\nano\metre}$, the screening length is $\SI{60}{\nano\metre}$, and the Fermi wavelength is comparable to the width of the narrow channel of $\SI{150}{\nano\metre}$.
Thus $\tilde{a}$ is the smaller length scale in our simulations by at least an order of magnitude.
Previous works kept values of $s$ to explore a high-magnetic field regime with $B\sim\SI{10}{\tesla}$, leading to $l_B\sim\SI{10}{\nano\metre}$~\cite{PhysRevLett.114.036601,Chen_2024}.

\section{Robustness to the splitter angle}
\label{app:angle}

\begin{figure}[h]
    \centering
    \includegraphics[width=\textwidth]{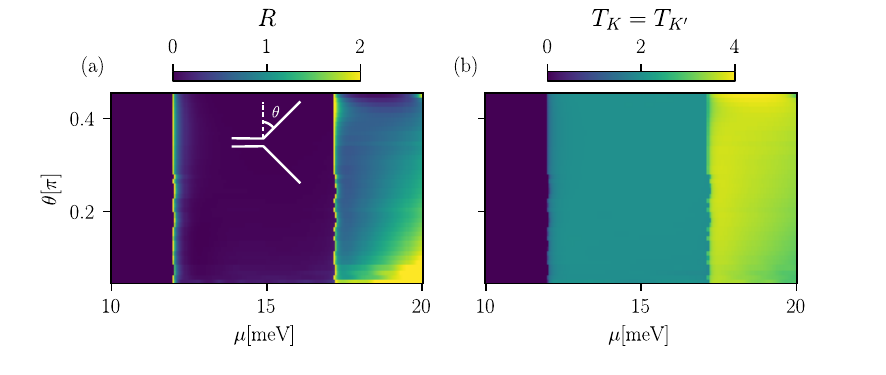}
    \caption{Reflection (a) and transmission (b) probabilities of injected modes as a function of the arm angle $\theta$ when the injected modes come from the lower QPC (see Fig.~\ref{fig:device} (c)).
 The transmission to the left and right leads, namely $T_K$ and $T_{K^\prime}$ is symmetric.
 The inset shows the angle $\theta$ between the helical channels.
 }
    \label{fig:angle}
\end{figure}

The propagation of helical edge states along the domain wall where the displacement field changes sign is topologically protected.
Thus, the transmission of helical edge states is independent of the underlying lattice orientation.
As a consequence, the device has perfect helical transmission regardless of the angle between the helical channels.
In Fig.~\ref{fig:angle}, we show the transmission (panel (a)) and reflection (panel (b)) probabilities of injected modes as a function of the arm angle $\theta$ (indicated as an inset in panel Fig.~\ref{fig:angle}(a)).
We observe that in the operational range ($T_K = T_{K^{\prime}} = 2$) the transmission and reflection are insensitive to $\theta$.
The operational range corresponds to the lowest band of the injecting channel.
For higher bands the transmission and reflection probabilities are not perfect, but the valley polarization is still preserved.
Furthermore, the reflection probability peaks at the bottom of the band because the velocities of injecting modes vanish.

\end{document}